%% file: saaoga.tex
\documentclass{aa}
\usepackage{graphics}

\input abb.tex

\begin{document}

\thesaurus{03(04.03.1; 
              04.19.1; 
              09.04.1; 
              11.04.1; 
              11.03.4; 
              12.12.1) 
                      }
\title 
{Extragalactic large-scale structures behind the southern Milky Way. 
-- III. Redshifts obtained at the SAAO in the Great Attractor region
\thanks
{All the tables also available in electronic form. See the 
Editorial in A\&AS 1994, Vol. 103, No.1. 
Based on observations taken at the South African Astronomical
Observatory.}
}

\author{Patrick A.~Woudt\inst{1} \and Ren\'ee C.~Kraan-Korteweg\inst{2} 
\and Anthony P.~Fairall\inst{3}}

\offprints {Patrick A.~Woudt (email: pwoudt@eso.org)}

\institute{
European Southern Observatory, Karl-Schwarzschildstr.~2, D-85748 Garching,
Germany
\and
Departemento de Astronom{\'{\i}}a, Universidad de Guanajuato, Apartado 
Postal 144, Guanajuato, Gto 36000, Mexico
\and
Department of Astronomy, University of Cape Town,
Rondebosch 7700, South Africa}

\date{5 March 1999 / 8 October 1999}
\titlerunning{Woudt, Kraan-Korteweg and Fairall: LSS behind the southern Milky Way}
\maketitle
\markboth{Woudt, Kraan-Korteweg and Fairall: Large-scale structures behind the southern Milky 
Way.}{III. Redshifts obtained at the SAAO in the Great Attractor region.}

\begin{abstract}

In the third of a series of papers on large-scale structures behind the 
southern Milky Way, we report here on redshifts obtained at the South
African Astronomical Observatory (SAAO) in the 
Great Attractor region ($318\deg \la \ell \la 340\deg, |b| \le 10\deg$, 
Woudt 1998).

This region encompasses the peak in the reconstructed mass density field,
associated with the Great Attractor (Kolatt \etal 1995, Dekel \etal 1998) and
covers the crossing of the Supergalactic Plane with the Galactic Plane.

Our deep optical galaxy search in the Zone of Avoidance (ZOA) in this region (Woudt 
1998) has resulted in the detection of 4423 galaxies with observed 
diameters larger than 0.2 arcmin. 
We have obtained reliable redshifts for 309 galaxies of the 4423 galaxies
with the ``Unit'' spectrograph (first with a Reticon, then with a CCD detector)
at the 1.9-m telescope of the SAAO. 
An additional 13 tentative redshifts are presented.
Before our survey, 127 galaxies had a previously recorded redshift (NED and 
SRC96). Given a small overlap with the literature (44 galaxies), we present 
here redshifts for 265 galaxies that had no previous recorded velocity.
In addition, we present central velocity dispersion ($\sigma_{\rm o}$) measurements
for 34 galaxies in ACO 3627.

It is known that the Great Attractor (GA) region is overdense in galaxies
at a redshift-distance of $v \sim 5000$ {\kms} (Fairall 1988, Dressler 1991, 
Visvanathan \& Yamada 1996, di Nella et al. 1997).
We realise here, however, that the Great Attractor region is dominated by 
ACO 3627 (hereafter referred to as the Norma cluster), a highly obscured,
nearby and massive cluster of galaxies close to the plane of the Milky Way 
($\ell$, $b$, $v$) = (325.3$\deg$, --7.2$\deg$, 4844 {\kms}) (Kraan-Korteweg 
et al.~1996, Woudt 1998). 

Previous redshift surveys in the GA region have failed to 
gauge the significance of the Norma cluster, primarily due to the 
diminishing effects of the Galactic foreground extinction 
on the partially obscured galaxies.
In the absence of the obscuring effects of the Milky Way, the Norma cluster
would have appeared as prominent as the well-known Coma cluster, but nearer
in redshift-space. This cluster most likely marks the bottom of the potential well
of the Great Attractor (Woudt 1998).

\end{abstract}


\section{Introduction}

In two previous papers, we reported on large-scale structures
behind the southern Milky Way in the Hydra--Antlia region (Kraan-Korteweg 
et al.~1995 - hereafter Paper I) and in the Crux region
(Fairall et al.~1998 - hereafter Paper II). In the present
paper, the third of this series, we present redshifts of galaxies obtained 
with the 1.9-m telescope of the SAAO 
in the Great Attractor (GA) region, a region adjacent to the Crux area 
(Paper II). 

Our deep optical galaxy search in the GA region has resulted in the detection 
of 4423 galaxies with major diameters larger than 0.2 arcmin. They were 
identified by visually scanning film copies of the SRC IIIaJ survey under 
50 times magnification.
Details of the galaxy search are given by Woudt (1998) and the results of 
the galaxy search will be presented as a catalogue (Woudt \& Kraan-Korteweg 
1999, hereafter WKK99, in prep.).  
The search in the GA region covers 16 fields of the SRC Sky Survey, namely
F99-100, F135-138, F176-180 and F221-225, covering $\sim$400 square degrees.

\begin{figure*}
 \resizebox{\hsize}{!}{\includegraphics{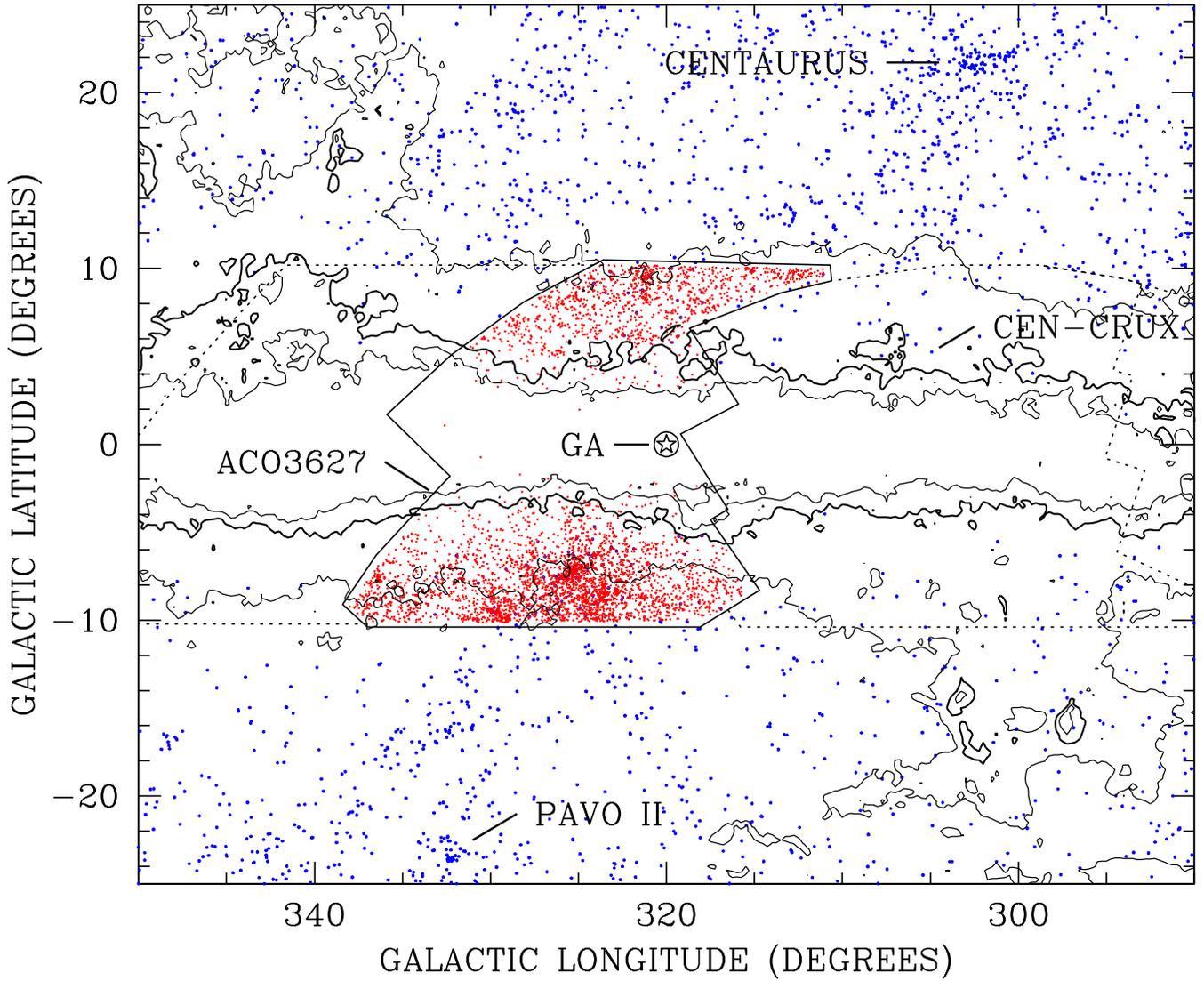}}
\caption
{The distribution of galaxies in the Great Attractor region. The solid
line outlines the surveyed area. The galaxies uncovered in the deep
optical search ($D \ge 0.2$ arcmin) are displayed as small dots.
The larger dots in the surrounding area are the Lauberts galaxies
($D \ge 1.0\arcmin$). The Centaurus, Pavo II, Centaurus-Crux and ACO3627
clusters are 
labelled, as is the location of the peak of the reconstructed mass density 
field associated with the Great Attractor. The contours are lines
of equal Galactic foreground extinction, taken from the Galactic 
reddening maps of Schlegel \etal (1998). The contours correspond to 
A$_{\rm B}$ = 1$^m$, 3$^m$ (thick line) and 5$^m$.
}
\label{saaof1}
\end{figure*}

The distribution in Galactic coordinates of the 4423 identified galaxies is 
shown in Fig.~1. A small fraction of these galaxies had been catalogued 
before by Lauberts (1982), namely 2.4\% (=108 galaxies). The adjacent Crux 
and Hydra--Antlia regions at lower Galactic longitude are demarcated by 
the dotted line in Fig.~\ref{saaof1}, as is the current extension towards
the Galactic Bulge (Fairall \& Kraan-Korteweg, in prep.).

By far the most prominent overdensity of galaxies in the GA region 
is centred on the nearby\footnote[1]{To avoid confusion about the 
redshift-distance of ACO 3627, we quote here the most reliable value 
as derived from the redshifts of 219 cluster members in the Abell 
radius (Woudt 1998).} ($cz = 4844$ {\kms}) Abell cluster ACO 3627 (Abell 
\etal 1989) at ($\ell, b) \approx (325\deg, -7\deg$) in the 
constellation of Norma. The galaxies in this overdensity are on average 
quite large ($<${D}$>$=30$\farcs$3) and bright ($<$B$_{\rm J}$$>$=16$\fm$9).
The large fraction of early type galaxies in ACO 3627 (50\% of the galaxies 
within the core radius are ellipticals or lenticulars) indicate that this is 
indeed a rich cluster of galaxies. Moreover, if corrected for the obscuring 
effects of the Galactic extinction (Cameron 1990), and only including galaxies
with extinction-corrected diameters D$_0 \ge 1\farcm{3}$,
this region would have the highest 
galaxy density in the entire southern sky (Woudt 1998).

This clearly suggests that we have unveiled a major
cluster of the nearby Universe. 
The central region of the Norma cluster is a factor $f = 8 - 10$ more dense 
compared to regions at similar Galactic latitude.
Due to the diminishing effects of the foreground extinction (A$_{\rm B} 
\approx 1^m - 2^m$), the richness of this centrally-condensed 
cluster had not previously been noticed, even though this cluster lies within 
$10\deg$ of the centre of the Great Attractor ($\ell, b) 
\approx (320\deg, 0\deg$), \cf Kolatt \etal (1995).

Another significant concentration of galaxies in the GA region is 
located not far from the Norma cluster, at ($\ell, b) \approx (329\deg, 
-9\deg$). The relatively large number of early-type galaxies in this 
overdensity are, on average, much smaller than the early-type galaxies in 
the Norma cluster, especially when taking into account that the Galactic 
foreground extinction is nearly identical for both clusters. This overdensity,
hereafter referred to as the Ara cluster, is possibly connected to the
X-ray bright, and distant Triangulum-Australis cluster 
at ($\ell, b, v$) = ($324\deg$, $-12\deg$, 15300 {\kms}) (McHardy \etal
1981). 

Hardly any galaxies are visible at extinction levels of A$_{\rm B}$ $\ge$ 5 
mag. Our deep optical galaxy search in the Great Attractor
region has reduced the optical `Zone of Avoi\-dance' to Galactic latitudes
$|\, b \, | \le 3\deg$.

\section {Observations}

As before in the Hydra--Antlia and Crux regions, we have aimed to 
be as complete as possible in tracing the bright end of the magnitude 
distribution of the identified galaxies in the GA region. In addition, 
particular
emphasis is given to the galaxies in the Norma cluster. Together with our
MEFOS (Meudon ESO Fibre Optical Spectrograph)
observations (Woudt et al. 1999, in prep.) we have aimed to observe
all the galaxies down to an extinction-corrected magnitude of 
${\rm B}_{\rm J}^{0} = 15\fm{5}$, see also Sect.~3.2.

The procedures used for observations at the SAAO are the same 
as described in Paper I. From the first quarter of 1997 onwards, a new 
CCD detector has
been installed at the 1.9-m telescope of the SAAO, replacing the old Reticon
Photon Counting System (RPCS). The RPCS data have been reduced at the
University of Cape Town, following the procedure outlined in Paper I. The
data obtained with the CCD camera have all been processed by using IRAF\footnote[1]{IRAF
(Image Reduction and Analysis Facility) is distributed by the National
Optical Astronomy Observatories, which are operated by the Association of
Universities for Research in Astronomy, Inc., under cooperative agreement
with the National Science Foundation} 
and
the various tasks within this software package, e.g. {\sl ccdred} and {\sl rvsao}.
Both the RPCS and CCD data were taken from March 1994 to May 1997
with grating 7 (210 {\AA}/mm), resulting in a wavelength coverage of
3500 -- 7000 {\AA}.

In Table~1\,\footnote[2]{All the tables are only available in 
electronic form at the CDS via anonymous ftp to cdsarc.u-strasbg.fr 
(130.79.128.5) or via http://cdsweb.u-strasbg.fr/Abstract.html}, 
the 322 galaxies for which we obtained a redshift
at the SAAO are listed. This table includes the 13 tentative redshifts,
and the galaxies observed at higher spectral resolution (see Sect.~2.1).
The entries in Table~1 are as follows:

\begin{description}

\item [Column 1 and 2:] Identification of the galaxy as given in WKK99
and Lauberts Identification (Lauberts, 1982). 

\item [Column 3 and 4:] Right Ascension and Declination (1950.0). The
positions were measured with  the Optronics machine at the 
ESO in Garching and have an accuracy of about 1 arcsec.

\item [Column 5 and 6:] Galactic longitude $\ell$
and latitude $b$.

\item [Column 7:] Major and minor axes (in arcsec). These diameters
are measured approximately to the isophote of 24.5 mag arcsec$^{-2}$ and
have a scatter of $\sigma \approx 4\arcsec$.

\item [Column 8:] Apparent magnitude B${\rm_J}$. The magnitudes are estimates
from the film copies of the SRC IIIaJ Survey based on the above given 
diameters and an estimate of the average surface brightness of the galaxy. 

\item [Column 9:] Morphological type. The morphological types are coded 
similarly to the precepts of the Second Reference Catalogue 
(de Vaucouleurs \etal 1976). Due to 
the varying foreground extinction a homogeneous and detailed type 
classification could not always be accomplished and some
codes were added: 
In the first column F for E/S0 was added to the normal designations
of E, L, S and I. In the fourth column the subtypes E, M and L are 
introduced next to the general subtypes 0 to 9. They stand for
early spiral (S0/a-Sab), middle spiral (Sb-Sd) and late spiral or 
irregular (Sdm-Im). The cruder subtypes are a direct indication of the 
fewer details visible in the obscured galaxy image. The 
question mark at the end marks uncertainty of the main type, the colon
uncertainty in the subtype.

\item [Column 10:] Heliocentric velocity (cz)
and error as derived from the absorption features. The errors may
appear large as they are estimated external errors, and not internal errors
(see Paper I). The square brackets indicate a tentative redshift.

\item [Column 11:] Heliocentric velocity and error measured
from the emission lines (identified in column 12) when present.
The square brackets indicate a tentative redshift.

\item [Column 12:] Identified emission lines: \hfill \phantom{a}

\smallskip

\begin{tabular}{ccccccc}
 1 & 2 & 3 & 4 & 5 & 6 & 7 \\
 $[$OII] & H$\gamma$ & H$\beta$ & [OIII] & [OIII] & H$\alpha$ & [NII] \\
 3727 & 4340 & 4861 & 4959 & 5007 & 6563 & 6584 \\
\end{tabular}
\end{description}

\begin{description}
\item [Column 13:] Code for additional remarks:\\

$*$ -- \hspace{0.15cm} Redshifts are also available in the literature.\\

1 -- \hspace{0.15cm} WKK 4001: The redshift measured at the SAAO for this 
galaxy ($v = 5937 \pm 250$ {\kms}) is in disagreement with the value quoted in 
the literature ($v = 4700 \pm 100$ {\kms}, Fairall 1981). \\


%

2 -- \hspace{0.15cm} WKK 5792: The redshift measured at the SAAO for this 
galaxy ($v = 13487 \pm 101$ {\kms}) is in disagreement with the value quoted 
in the literature ($v = 3500 \pm 70$ {\kms}, di Nella et al. 1997). There is a 
positional mismatch of 1.1 arcminutes between WKK 5792 and the galaxy 
quoted by di Nella et al. (1997). However, after visual re-examination, 
no galaxy is seen on the IIIaJ Sky Survey at the position of di Nella's 
galaxy and we believe that this velocity should not be trusted.\\

3 -- \hspace{0.15cm} WKK 6180 / 6212: The redshifts measured for WKK 6180 
and WKK 6212 ($v = 4619 \pm 127$ {\kms} and $3196 \pm 95$ {\kms}, respectively)
are in disagreement with the value quoted in the literature 
($v = 3100$ {\kms} and $v = 4911$ {\kms}, respectively, West \etal 1981). 
This might be due to an identification error. \\

4 -- \hspace{0.15cm} WKK 7055: The spectrum is contaminated by the light
of a Galactic foreground star. However, a reliable redshift could be obtained.\\

GR6 -- These galaxies have been observed at higher spectral resolution with
grating 6 (100 {\AA}/mm), see also Sect.~2.1.\\

Sy1 -- This galaxy, WKK 6092, has been classified as Seyfert 1 (Woudt \etal
1998).\\

\end{description}

Table~2\,\footnote[1]{All the tables are only available in electronic form 
at the CDS via anonymous ftp to cdsarc.u-strasbg.fr (130.79.128.5) or via 
http://cdsweb.u-strasbg.fr/Abstract.html} 
lists 22 galaxies for which no redshift could be determined. These 
spectra either had a poor signal-to-noise ratio or were dominated by the 
light of a Galactic foreground star.

Table~3\,\footnotemark[1] 
gives redshifts extracted from the literature for 82 galaxies we have not
observed in the Great Attractor region. 
These galaxies would have been included in our observations were they not
observed before, since they meet our selection criteria.

Columns 1-9 are the same as in Table~1. 
Column 10 list the heliocentric
velocities and errors (if given). The velocity
in column 10 has been adopted from the source identified in
column 11, where the number corresponds to:

\begin{enumerate}
\item{Dressler 1991}
\item{di Nella et al. 1997}
\item{Fisher et al. 1995}
\item{Cot\'e et al. 1997}
\item{Strauss et al. 1992}
\item{Jones \& McAdam 1992}
\item{Huchtmeier \& Richter 1989}
\item{Webster 1979}
\item{Visvanathan \& Yamada 1996}
\item{Mould et al. 1991}
\item{Fairall 1988}
\item{Davies et al. 1989}
\item{Fairall 1983}
\item{Schmidt \& Boller 1992}
\item{Fairall 1981}
\item{Whiteoak \& Gardner 1977}
\item{Corwin \& Emerson, 1982}
\end{enumerate}

\subsection{Velocity dispersion measurements for galaxies in the Norma
cluster}

During one week in April 1996, we observed 39 galaxies in the Norma
cluster at higher spectral resolution, \ie with grating 6 (100 {\AA}/mm
= 1.4 {\AA}/pixel) in combination with the RPCS. 
These galaxies had been observed before, 
either by us in the course of our redshift survey (34 galaxies), 
or by others (5 galaxies).
The higher spectral resolution allows the determination of the 
central velocity dispersion of these galaxies.
The wavelength coverage ranges from 4100 {\AA} to 6050 {\AA}, i.e. 
approximately centred on the Mg$\, b$ absorption lines at $\lambda_0$ = 
5172 {\AA}.
The slit width is 1.8 arcsec and the length of each slit segment is 6 arcsec.
The spectral resolution is approximately 2.9 {\AA}. This was measured 
from the width of the lines of the calibration lamp, and from the width 
of the peak of the autocorrelation function of the velocity standard stars.
The typical exposure times range between 1500-s and 2000-s per galaxy.

Three standard stars of spectral type G8 to K2 were observed
to serve as templates for the velocity dispersion determination.
We have used the Tonry \& Davis (1979) cross-correlation technique,
implemented in the IRAF task {\sl fxcor}, to determine the central
velocity dispersion. Scodeggio \etal (1998) reported that {\sl fxcor}
overestimates the velocity dispersion at a 5\% level, so a small correction
needs to be applied to the measured velocity dispersion.

One galaxy, WKK6080, only revealed [OIII] and H$\beta$ in emission, 
and hence no velocity dispersion measurement could be made for this galaxy.
For four further spectra (WKK5919, WKK6221, WKK6298 and WKK6580) no reliable
velocity dispersion could be determined because of low signal-to-noise
ratios.

Table~4\,\footnotemark[1] lists the final 34 galaxies and their
respective central velocity dispersion. The entries in Table 1 are:

\begin{description}

\item [Column 1 -- 8:] As in Table 1.

\item [Column 9:] The Galactic reddening at the position of each galaxy,
taken from the Galactic reddening maps of Schlegel \etal (1998).

\item [Column 10:] As Column 9 of Table 1.

\item [Column 11:] As Column 10 of Table 1.

\item [Column 12:] The measured central velocity dispersion. In brackets,
the number of standard stars are given that were used for the 
determination of the velocity dispersion. A correction is applied to the
velocity dispersion following Scodeggio \etal (1998), since the IRAF task
{\sl fxcor} overestimates the velocity dispersion at a 5\% level.

\item [Column 13:] The logarithm of the aperture-corrected (Davies \etal 1987)
central velocity dispersion, and its uncertainty calculated from the signal-to-noise
ratio of the spectra (Scodeggio \etal 1998). 

\end{description}

\subsection{Comparison to other measurements}

In the course of this redshift survey, some galaxies in our search area
have been observed spectroscopically by others (Visvanathan \& Yamada 
1996, di Nella \etal 1997). Although the galaxies we observed were 
initially selected on the basis of having no published redshifts, a small 
overlap now exists.

This overlap allows for a comparison between our sample and others.  
We find
\begin{displaymath}
<v_{\rm SAAO} - v_{\rm pub}> = +38 \pm 151 {\rm \ km \,\, s^{-1}}.
\end{displaymath}  
 
which shows no significant systematic error, and agrees well 
within our average standard deviation. This is based on 43 galaxies for
which a redshift estimate exists in the literature (not including
WKK 4001 and WKK 5792).
 
Similarly, we have allowed a small overlap between the SAAO 
galaxies and our complementary programmes using MEFOS and Parkes radio 
observations (Kraan-Korteweg \etal 1994), for which we find

\begin{displaymath}
<v_{\rm SAAO} - v_{\rm MEFOS}> = +3 \pm 148 {\rm \ km \,\, s^{-1}},
\end{displaymath}
\begin{displaymath}
<v_{\rm SAAO} - v_{\rm Parkes}> = -10 \pm 238 {\rm \ km \,\, s^{-1}}.
\end{displaymath}
 
The comparison with the MEFOS and Parkes data is based on 24 and 8 galaxies, 
respectively. The agreement is good.
The comparison with Parkes redshifts are mainly for low surface-brightness
galaxies, for which our errors are larger.

Note that in some cases, the difference between the optical and HI-based 
radial velocity is real due to the net 
outflow of gas in the narrow emission line
regions, as for instance in the case of Seyfert galaxies (Mirabel \& Wilson
1984). One such galaxy, a Seyfert 1 in ACO 3627 (WKK 6092, see Woudt \etal 
1998), has been included in the above statistics. If this galaxy is excluded, 
we find

\begin{displaymath}
<v_{\rm SAAO} - v_{\rm Parkes}> = +35 \pm 218 {\rm \ km \,\, s^{-1}}.
\end{displaymath}

A final comparison is made for galaxies reobserved at the SAAO at
higher spectral resolution with grating 6. All these galaxies
are bona fide members of the Norma cluster.
Eleven galaxies were observed before with MEFOS on the 3.6-m telescope of
ESO, La Silla. For these galaxies we find no significant offset, and a very
small standard deviation which reflects the high signal-to-noise ratio
at higher spectral resolution for their velocity dispersion determination.

\begin{displaymath}
<v_{\rm SAAO, \ gr.6} - v_{\rm MEFOS}> = +17 \pm 49 {\rm \ km \,\, s^{-1}}.
\end{displaymath}

For 23 of these galaxies, both a low and high spectral resolution spectrum
are available from our SAAO data. For these galaxies we find

\begin{displaymath}
<v_{\rm SAAO, \ gr.6} - v_{\rm SAAO}> = +28 \pm 136 {\rm \ km \,\, s^{-1}}.
\end{displaymath}

There are only a few elliptical and lenticular galaxies in the Zone of 
Avoidance for which the central velocity dispersion has been measured.
Dressler \etal (1991) presented data for the three brightest galaxies in
the Norma cluster. These galaxies (WKK6269 -- the central cD galaxy, 
WKK6312 -- another cD galaxy, and WKK6318 -- a bright elliptical galaxy) 
are also part of our sample of 34 galaxies.

It is difficult to make a quantitative comparison based on this small overlap.
Moreover, the three galaxies in common are biased in the sense that
they are the brightest ellipticals in the Norma cluster.
For these three galaxies we find

\begin{displaymath}
<\log{({\sigma_{\rm o})}_{\rm SAAO}} - \log{({\sigma_{\rm o})}_{\rm lit}}> = +0.109 \pm 0.074.
\end{displaymath}

Note that one galaxy (WKK6269) is in perfect agreement with Dressler \etal (1991), but
that WKK6312 and WKK6318 have a larger dispersion compared to those presented by
Dressler \etal (1991). Although we measure a larger velocity dispersion for WKK6312
and WKK6318, these are not atypical values for a cD galaxy, and a bright elliptical galaxy, 
respectively.

A very recent paper by Lucey et al. (1999, in prep.) reports on velocity dispersion
measurements of galaxies in ACO 3627. For six galaxies with sufficiently high signal-to-noise
ratio we find a good agreement between the two data sets,

\begin{displaymath}
<\log{({\sigma_{\rm o})}_{\rm SAAO}} - \log{({\sigma_{\rm o})}_{\rm Lucey}}> = -0.002 \pm 0.085.
\end{displaymath}

\begin{figure}
 \resizebox{\hsize}{!}{\includegraphics{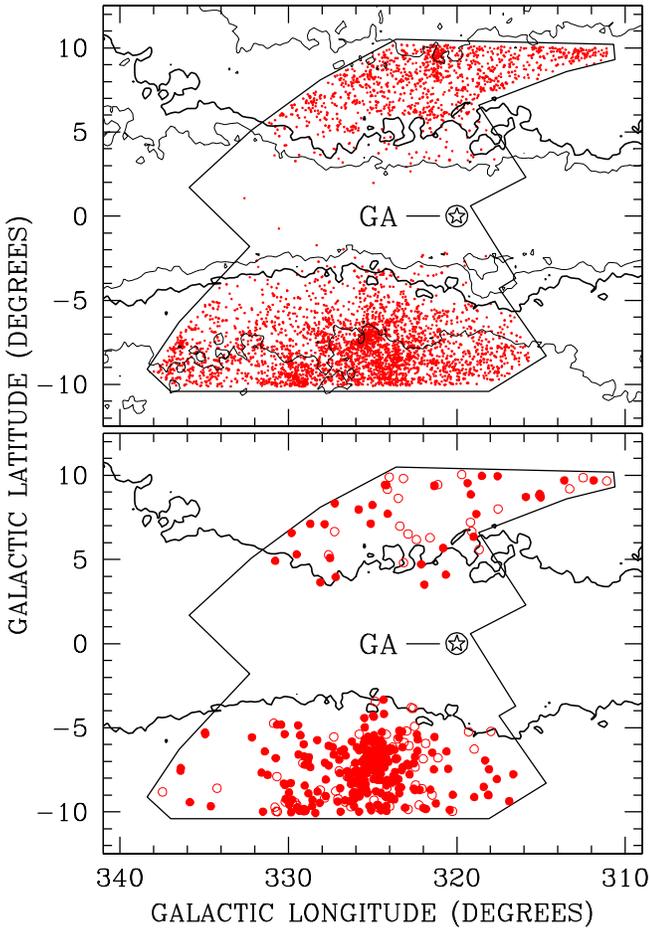}}
\caption {The top panel shows the distribution in Galactic coordinates
of the partially obscured galaxies in the Great Attractor region.
The displayed 
contours mark the foreground extinction (Schlegel \etal 1998), see Fig.~1. 
The thick line (also shown in the bottom panel) 
corresponds to ${\rm A_B} = 3\fm{0}$, below this line our deep optical 
galaxy survey is not complete anymore for galaxies with ${\rm D^0} \ge 
1\farcm$3. The bottom panel shows the distribution of the galaxies with 
radial velocities. The GA survey region is outlined.
Solid circles indicate the positions of galaxies, in the GA region, 
observed in the present work, and for which redshifts have been obtained.  
Open circles show the positions of galaxies for which redshifts are 
available from the literature. }
\label{saaof2}
\end{figure}

\section{Coverage and Completeness}

\subsection{The GA region}
 
The top panel in Fig.~\ref{saaof2} shows the distribution in
Galactic coordinates of the newly identified galaxies in the GA region.
The contours mark the Galactic foreground extinction (Schlegel \etal 1998).
The bottom panel of Fig.~\ref{saaof2} shows the sky coverage
of galaxies with reliable redshifts, indicated by the solid circles.
The open circles correspond to galaxies with previously known redshifts.
It is clear that a large fraction of our observational effort was 
directed towards obtaining a fairly complete coverage of the Norma cluster.

\subsection{ACO 3627: The Norma cluster}

\begin{figure}
 \resizebox{\hsize}{!}{\includegraphics{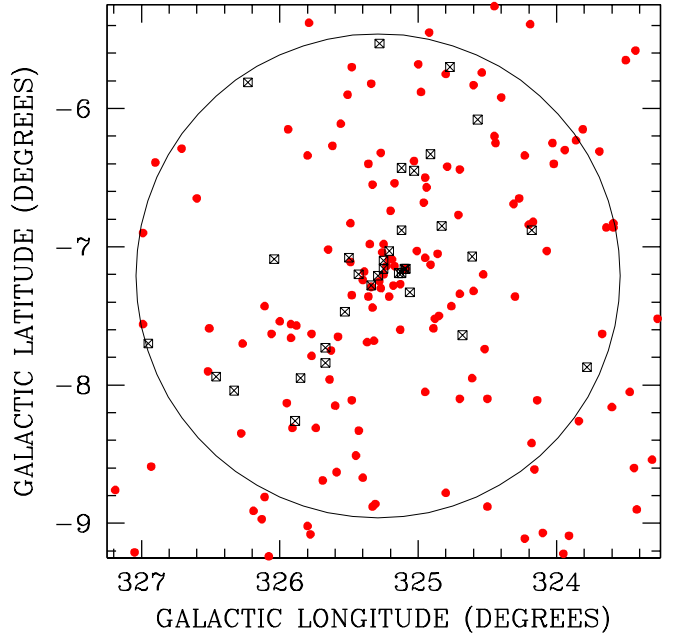}}
\caption {The distribution of the galaxies with radial velocities in the
immediate vicinity of the Norma cluster. 
The crossed squares are the
galaxies for which we have obtained a central velocity dispersion. The filled circles 
correspond to the galaxies with a reliable redshift (both the newly obtained SAAO
redshifts as well as the literature data).
The circle marks the Abell radius of the Norma
cluster (3 ${\rm h}_{50}^{-1}$ Mpc, or 1.75 degrees on the sky at the 
redshift-distance of the Norma cluster). }
\label{saaof2b}
\end{figure}

A more detailed view of the sky coverage of galaxies with reliable 
redshifts in the immediate vicinity of the Norma cluster is shown in
Fig.~\ref{saaof2b}. 

The filled circles in Fig.~\ref{saaof2b} correspond to all the galaxies
with a known redshift (SAAO and literature data), whereas
the crossed squares in Fig.~\ref{saaof2b} show the 34 galaxies for which
we have obtained the central velocity dispersion. The latter sample is spread
fairly uniformly over the entire cluster and represents a fair subsample of
the Norma cluster. 
It has a mean velocity of 4877 {\kms} and a dispersion of 791 {\kms}, compared
to 4844 $\pm$ 63 {\kms} and 848 {\kms} of all Norma cluster members (Woudt 1998).

Apart from our standard observing strategy, \ie to obtain a spatially
uniform coverage of the bright end of the luminosity distribution of the
newly catalogued galaxies in the GA region (see also papers I and II), 
we have aimed to observe all the galaxies brighter than 
${\rm B}_{\rm J}^{0} = 15\fm{5}$ within the Abell radius of the 
Norma cluster. 

The top panels of Fig.~\ref{compl} show the magnitude and major-axis
distribution of all the galaxies in the Great Attractor region
with a reliable redshift. The SAAO data are indicated by the lighter hatched
histogram, the dark shaded histogram shows the literature data.

We have achieved a similar completeness compared to the Hydra--Antlia and Crux
regions {Paper I and II). We are
91\% complete for galaxies brighter than (B$_{\rm J} \le 14^m$)
and even 56\% complete for galaxies brighter than (B$_{\rm J} \le 16^m$).
	
The lower panels of Fig.~\ref{compl} show the {\sl extinction-corrected}
magnitude and major-axis distribution of the 603 galaxies within the Abell
radius of the Norma cluster (open histogram).
Together with data from our MEFOS and Parkes redshift survey, and with 
literature data, we have now obtained a reliable redshift for 83\% of the 
galaxies with ${\rm B}_{\rm J}^{0} = 15\fm{5}$ within 3 Mpc of the Norma cluster.
This is illustrated by the cross-hatched histogram in the lower panels of
Fig.~\ref{compl}. For the remaining 36 galaxies (=17\%) brighter than
${\rm B}_{\rm J}^{0} = 15\fm{5}$ without redshifts, we
have observed a further 10 galaxies, but no reliable redshift could be obtained.
For galaxies brighter than ${\rm B}_{\rm J}^{0} = 14\fm{5}$, 95\% of the 
galaxies were observed and a reliable redshift has been obtained for 91\%.

The redshifts obtained at the SAAO (dark shaded histogram in the lower
panels of Fig.~\ref{compl}) account for more than 50\% of the
newly obtained redshifts within the Norma cluster.

\begin{figure}
 \resizebox{\hsize}{!}{\includegraphics{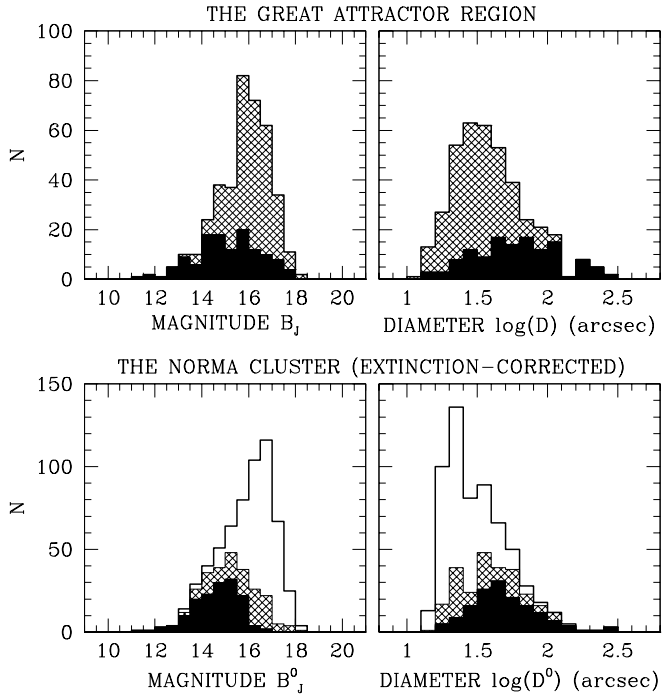}}
\caption {The top panel shows the magnitude and major-axis distribution
of galaxies with radial velocities in the GA region. The lighter hatched
areas mark the galaxies observed at the SAAO, the dark shaded histogram 
are galaxies previously observed by others. The lower panel shows the
extinction-corrected magnitude and major-axis distribution of the galaxies
in the Norma cluster. The open histograms mark all the optically identified
galaxies, the cross-hatched histograms illustrate those galaxies that now
have a reliable redshift (including MEFOS, Parkes and literature data) and
the dark shaded histograms show the redshifts presented in this paper.}
\label{compl}
\end{figure}

\section {Identification of Large-Scale Structures}

\subsection{Velocity distribution}

Fig.~\ref{saaof4} shows the redshift distribution of the observed 
galaxies in the GA region. 
The literature data reveal shallow peaks at 3000 {\kms} and 5000 {\kms}.
With the new redshifts, the peak at 5000 {\kms} is the most dominant 
feature in the GA region. Most of the galaxies at this redshift-distance 
belong to the Norma cluster (see Fig.~\ref{saaof4b}), and its immediate 
surrounding. A more distant (broad) peak can be seen at 15000 {\kms}. 
This peak is associated with the Ara cluster.

In Fig.~\ref{saaof4b}, only the galaxies within the Abell radius of
the Norma cluster (the circle in Fig.~\ref{saaof2b}) are plotted.
From this velocity histogram it is clear that

\begin{itemize}
\item{The strong peak at 5000 {\kms} in Fig.~\ref{saaof4} is primarily
due to the Norma cluster.}
\item{The Norma cluster has a large velocity dispersion of $\sigma = 845$ 
{\kms}. This large velocity dispersion is indicative of a very massive
cluster. According to the 
Virial theorem and following Sarazin (1986), it translates into a 
dynamical mass of $1 \times 10^{15}$ M$_{\odot}$
within a 3 h$_{50}^{-1}$ Mpc radius (see also Kraan-Korteweg \etal 1996).
The dynamical mass could not have been determined on the basis of the 
available literature data (20 galaxies, dark shaded histogram in Fig.~\ref{saaof4b}).}
\item{Fore- and background galaxies are clearly offset from the main
cluster members. There is hardly any contamination by field galaxies
at the cluster core. There are 152 redshifts known within the
Abell radius of the Norma cluster, and 137 galaxies (=90\%) are
cluster members.}
\end{itemize}

Therefore, from the SAAO data alone, one can conclude that
the Norma cluster is a massive cluster near the heart of the
Great Attractor.

\begin{figure}
 \resizebox{\hsize}{!}{\includegraphics{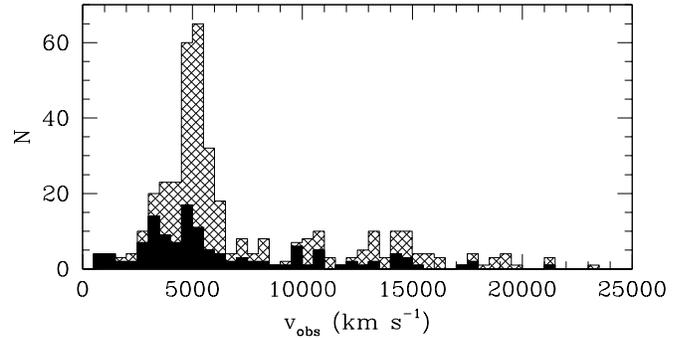}}
\caption {Velocity histogram of the galaxies in the search area
in the Great Attractor extension in the ZOA. Lighter hatched areas are 
velocities measured by us; darker hatched by others.}
\label{saaof4}
\end{figure}

\begin{figure}
 \resizebox{\hsize}{!}{\includegraphics{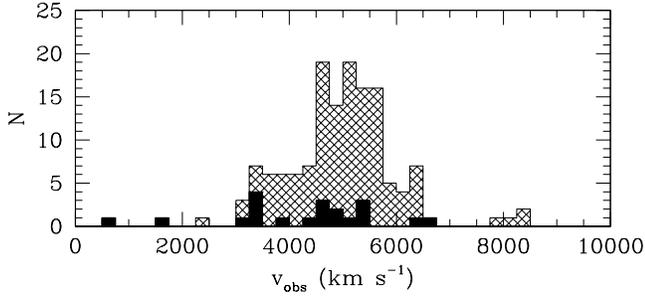}}
\caption {Velocity histogram of the galaxies within the Abell radius
of the Norma cluster. Lighter hatched areas are velocities
measured by us at the SAAO; darker hatched by others.}
\label{saaof4b}
\end{figure}

\subsection{Sky projection}

Before examining the redshift distribution in the sky projection shown here, 
the reader should be aware that the galaxies plotted in any of the 
subsequent diagrams constitute an `uncontrolled' sample of galaxies. 
The complementary data adjacent to our survey region are taken from the 1996 
Southern Redshift Catalogue (SRC) (Fairall 1996) which 
lists galaxies purely on the basis of them having redshifts.

In Fig.~\ref{saaof5}, we have plotted the galaxy distribution
in Galactic coordinates sliced in redshift intervals of 
$\Delta v = 2000$ {\kms} out to 16000 {\kms}.
The most conspicuous features occur in the second, third and eight slice, 
coinciding with the peaks observed in the velocity histograms 
(see Fig.~\ref{saaof4}). This figure covers the same region of the sky 
as Fig.~\ref{saaof1}, where the major concentrations of galaxies are labelled.

In the first slice ($v \le 2000$ {\kms}), the 
Supergalactic Plane dominates the graph.
Our GA search region fully covers
the crossing of the Supergalactic Plane with the Galactic Plane. 
An excess of galaxies is seen within the Supergalactic Plane near 
the Centaurus A group (at $\ell = 310\deg$ and $b = 19\deg$). 
At higher Galactic longitudes ($340\deg \le \ell \le 360\deg$), the Local Void 
is quite distinct.

In the second slice ($2000 \le v \le 4000$ {\kms}), the new
data clearly suggest the presence of a narrow filamentary structure 
running from $\ell = 340\deg$, $b = -25\deg$ to the Centaurus cluster at 
$\ell = 303\deg$, $b = 20\deg$. In Paper II, we already noticed the extension
of this structure at low Galactic latitude northwards of the Galactic Plane.
Here, we can clearly identify the continuation of this structure below
the Galactic Plane, from where it continues to $(\ell, b)$ = ($340{\deg}, -25\deg$).
This extended overdensity is part of a Great Wall-like structure seen edge-on -- the 
Centaurus Wall (Fairall 1998). 

The Norma cluster becomes very pronounced in the third redshift slice 
($4000 \le v \le 6000$ {\kms}).
This redshift slice corresponds to the redshift-distance 
of the Great Attractor overdensity, \ie $\sim 4500$ {\kms} (Lynden-Bell 
\etal 1988, Kolatt \etal 1995)
and coincides with the strong single peak seen in the velocity 
histogram of the GA region (Fig.~\ref{saaof4}).
The new data, together with the neighbouring galaxies outside the survey area 
in an extended region around the Pavo II cluster ($\ell \approx 332\deg$, 
$b \approx -23\deg$) suggest a broad large-scale structure running more or less 
horizontally across the diagram. There is no clear connection to the 
Centaurus cluster which is located a lower redshift-distance (\eg Stein 
\etal 1997 and references therein). 
This broad feature, also referred to as the ``Norma supercluster'', was already
noted in Paper II. 

Traces can also be seen in the following slice, so the feature is also 
probably wall-like seen roughly side on -- \ie its width (or depth in 
Fig.~\ref{saaof5}) being some 3000 {\kms}, and its thickness several 
hundred {\kms} unless much is still hidden by the dense obscuration. 

The Norma supercluster is located at a greater distance compared to 
the Centaurus Wall mentioned above, but must be similarly massive. It includes
a cluster/group of galaxies around $(\ell, b, v)$ = ($305.5{\deg}, 
+5.5{\deg}, 6214$ {\kms}) (the Centaurus--Crux cluster, Paper II). 
The Centaurus--Crux cluster and the Vela overdensity (Kraan-Korteweg \& Woudt 
1993) at $(\ell, b, v)$ = $(280{\deg}, +6{\deg}, 6000$ {\kms}) 
probably form part of the Norma supercluster.

It is likely that the flow motions that led to the prediction of a Great
Attractor originate in fact from the confluence of these two massive
structures (the Norma supercluster and the Centaurus Wall), at the intersection
of which resides the massive Norma cluster.

Although the rich Norma cluster seems to constitute the bottom of the potential
well of the GA, it cannot be excluded that other major features of the
GA remain hidden by the Galactic foreground extinction. 
There are various indications that the strong extragalactic radio source
PKS1343-601, located at ($\ell, b, v$) = ($309.7{\deg}, +1.7{\deg}, 3872$
{\kms}), could mark the centre of a highly obscured (A$_{\rm B}$ = 12$^m$)
rich cluster in the GA overdensity (Woudt 1998, Kraan-Korteweg \& Woudt 1999).
The prospective PKS1343-601 cluster is located in the second
slice of Fig.~\ref{saaof5}, near to the intersection of the
Norma supercluster and the Centaurus Wall.

In the last redshift slice ($14000 \le v \le 16000$ {\kms})
the Ara cluster is seen at ($\ell, b$) = ($329\deg, -9\deg$). 
This cluster is located near the X-ray bright Triangulum--Australis 
cluster at ($\ell, b, v$) = ($324\deg, -12\deg, 15300$ {\kms}) (McHardy \etal 
1981). Together, they might be part a larger structure, \ie a supercluster.

\begin{figure*}
\hspace{1.5cm} \resizebox{15cm}{!}{\includegraphics{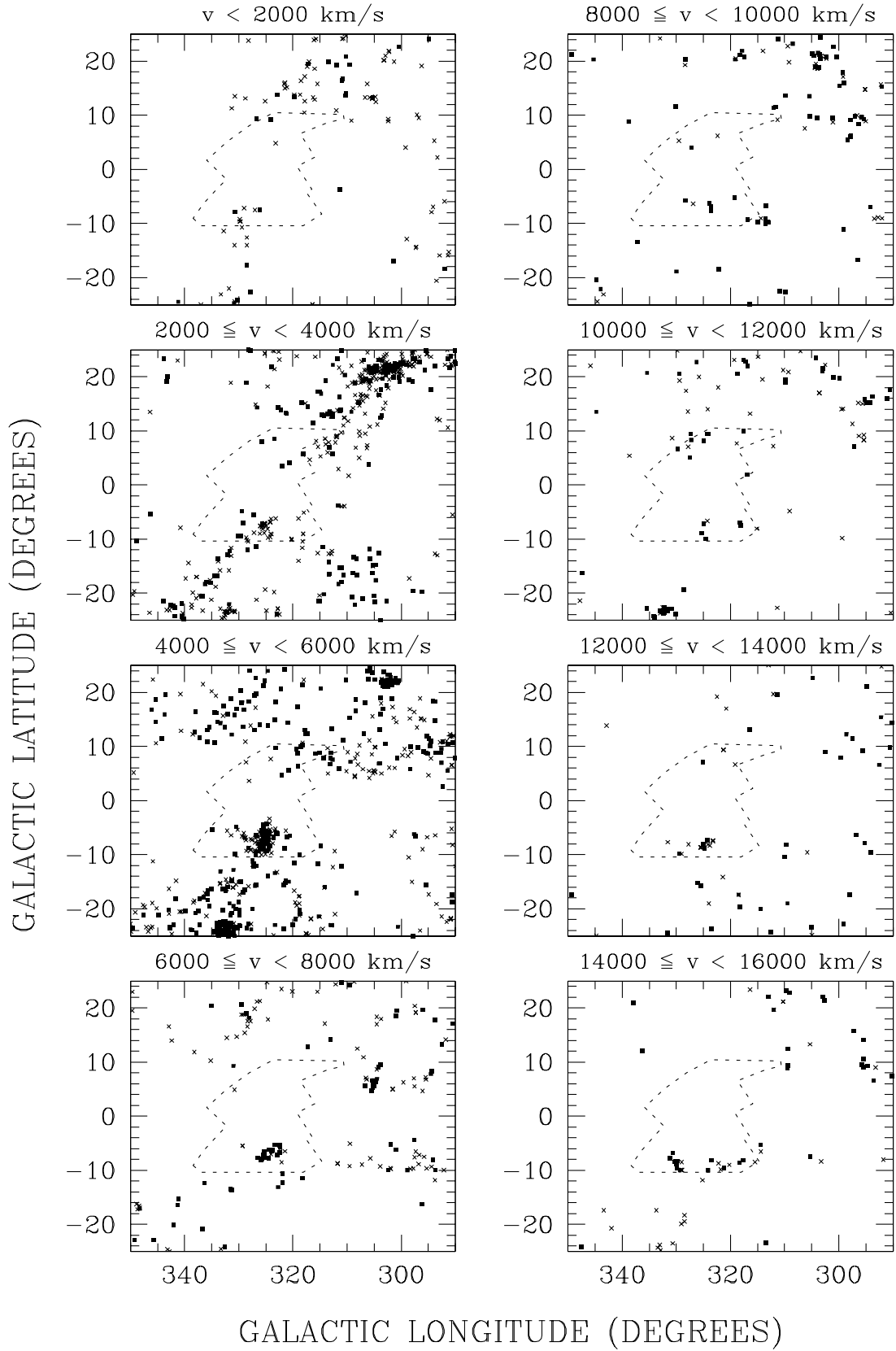}}
\caption{Sky projections in Galactic coordinates for redshift intervals
of $\Delta v=2000$ {\kms}. Within the panels the redshifts are subdivided
into intervals of $\Delta v=1000$ {\kms}: filled squares mark the nearer
redshift interval (\eg $v < 1000$ {\kms} in the top-left panel),
crosses the more distant interval ($1000\le v<2000$ {\kms} in same panel).
The sky plots increase in velocity-distance from the top-left panel to 
the bottom-right panel as marked above each panel.
The area of our investigation is outlined.}
\label{saaof5}
\end{figure*}

\begin{figure*}
 \resizebox{\hsize}{!}{\includegraphics{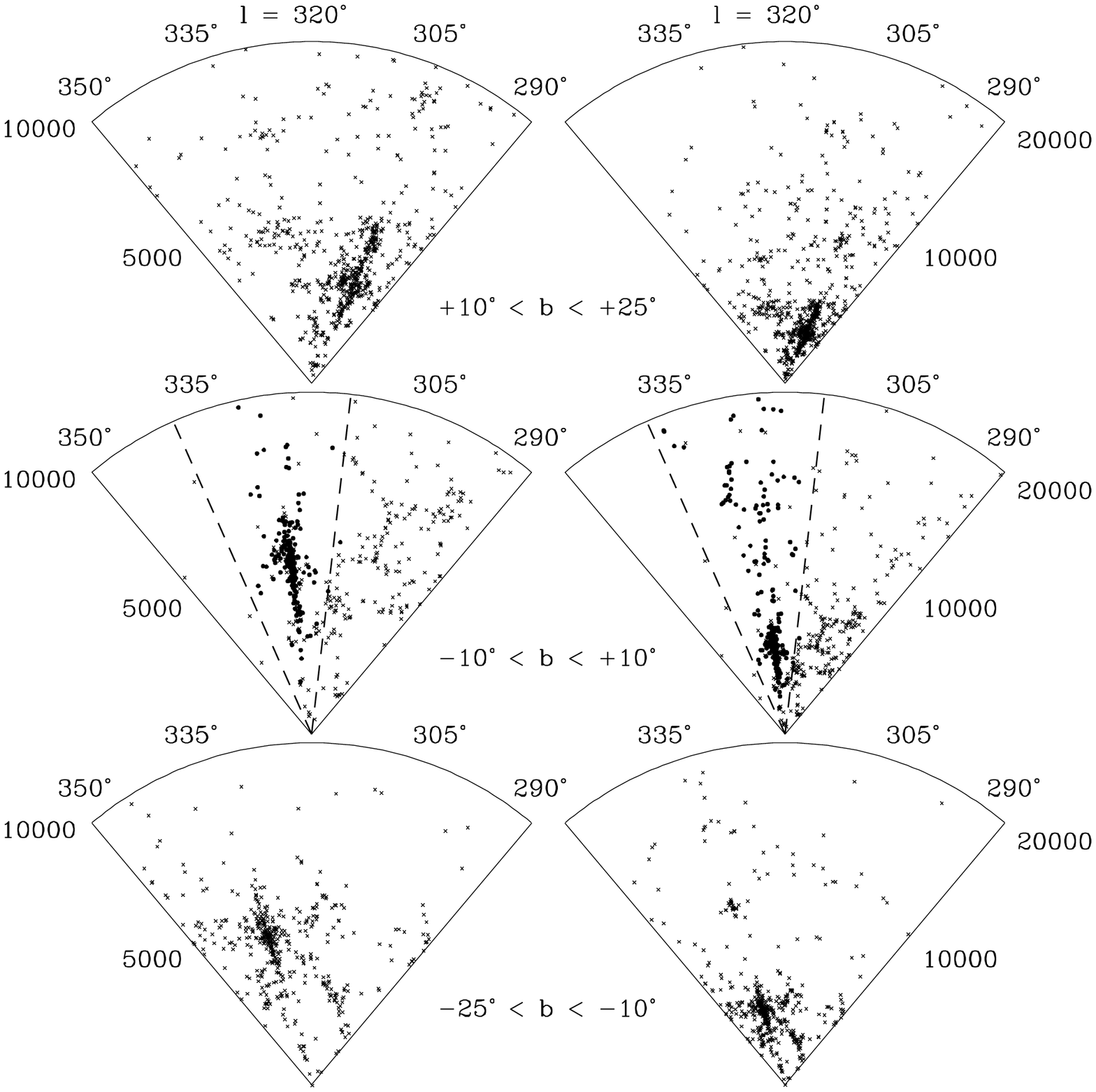}}
\caption{Redshift slices out to $v < 10000$ {\kms} (left panel) and
 $v < 20000$ {\kms} (right panel) for the longitude range $290\deg < \ell <
350\deg$. The top panels display the structures above
the GP ($+10\deg < b \le +25\deg$) the middle panel in the GP 
($-10\deg \le b \le +10\deg$) and the bottom panel the structures below
the GP ($-25\deg < b \le -10\deg$). The dashed lines in the middle panel
delimits the survey area. Filled squares are measurements from
the SAAO, crosses from the literature.}
\label{saaof6}
\end{figure*}

\subsection{Pie diagrams}

The redshift cones of Fig.~\ref{saaof6}
clearly show the impact of our redshift survey on the ZOA. The middle 
panels in Fig.~\ref{saaof6} mark the latitude interval 
$-10\deg \le b \le 10\deg$, a region that previously was largely 
blank now shows clusters, superclusters and voids. In this representation,
the ZOA out to $\ell = 340\deg$ is 
indiscernible from its unobscured counterpart.

The left panel of Fig.~\ref{saaof6} shows the galaxies out to a redshift 
of 10000 {\kms}, whereas the right panel shows the galaxies out to twice 
that distance. The upper panel includes the Centaurus clusters 
(Cen30 and Cen45), the lower panel includes the Pavo II cluster.
The Norma cluster, at $\ell = 325\deg$ in the ZOA, is very radially extended, 
indicative of the massive nature of this cluster. 
The Centaurus--Crux cluster at $\ell = 305\deg$ (Paper II) 
appears as a smaller finger of god. 
Several voids can be identified. One of the larger voids in the ZOA 
is located at $\ell \approx 316\deg, v \approx 7000$ \kms and has a radius
of $R \approx 1000$ {\kms} is the Circinus void (Fairall 1998). 
It probably connects with the void located behind the GA
(da Costa \etal 1996). Further nearby voids can easily be recognised.

\section {Summary}

Ever since the discovery of the Great Attractor -- a massive overdensity 
partly responsible for the large-scale systematic flow of galaxies in the 
Local Universe
-- it was clear that a large fraction of this overdensity was
hidden from our view by the obscuring veil of the Milky Way.

Our ZOA redshift survey at the SAAO has resulted in 265 new reliable 
redshifts in the Great Attractor region. These data clearly show that the
Great Attractor region is dominated by a rich and massive 
cluster at low Galactic latitude, \ie the Norma cluster. 

Moreover, it now emerges that the Great Attractor itself is likely to be 
the confluence of two massive 
large-scale structures in this part of the sky, the Centaurus Wall and 
the partially obscured Norma supercluster.

\acknowledgements

{The authors would like to thank the night assistants Francois van Wyk
and Fred Marang as well as the staff at the SAAO for their hospitality. 
We kindly acknowledge John Lucey, who made velocity dispersion measurements of 
16 early type galaxies in ACO 3627 available to us before publication.
Bruce Bassett assisted APF during the March 1994 observing run.
APF is supported by the South African FRD. This research has made use of 
the NASA/IPAC Extragalactic Database (NED), which is operated by the Jet 
Propulsion Laboratory, Caltech, under contract with the National Aeronautics 
and Space Administration.
}

\input H1430tab123.tex

\input H1430tab4.tex

\end{document}

%% file: abb.tex
\def\la{\mathrel{\hbox{\rlap{\hbox{\lower4pt\hbox{$\sim$}}}\hbox{$<$}}}}
\def\ga{\mathrel{\hbox{\rlap{\hbox{\lower4pt\hbox{$\sim$}}}\hbox{$>$}}}}

\def\arcmin{\hbox{$^\prime$}}
\def\arcsec{\hbox{$^{\prime\prime}$}}

\def\fm{\hbox{$.\!\!^{m}$}}

\def\farcm{\hbox{$.\mkern-4mu^\prime$}}
\def\farcs{\hbox{$.\!\!^{\prime\prime}$}}

\newcommand{\etal}{{et al.}\,}      
\newcommand{\eg}{{\it e.g.},\ }         
\newcommand{\ie}{{\it i.e.},\ }         
\newcommand{\cf}{{\it cf.},\ }          
\def\deg{{^\circ}}

\newcommand{\kms}{{\,km\ s$^{-1}$\,}}

%% file: H1430tab123.tex
\begin{table*}
 \normalsize
 \renewcommand{\baselinestretch}{0.75}
\caption{Redshifts of partially obscured galaxies in the Great Attractor region obtained at the SAAO.}
\label{tab1}
\scriptsize
\begin{tabular*}{18cm}{ 
  r @{\extracolsep{2mm}} p{1.5cm} @{\extracolsep{2mm}} 
  l@{\extracolsep{3mm}} l @{\extracolsep{4mm}}
  r @{\extracolsep{2mm}}r @{\extracolsep{2mm}} 
  p{6.5mm} @{\extracolsep{0mm}} p{4.5mm} @{\extracolsep{2.5mm}}
  r @{\extracolsep{3mm}} 
  p{2.7mm} @{\extracolsep{-1mm}} p{2.7mm} @{\extracolsep{-1mm}} 
  p{2.7mm} @{\extracolsep{-1mm}} 
  p{2.7mm} @{\extracolsep{-0.5mm}} p{2.7mm} @{\extracolsep{2mm}} 
 r @{\extracolsep{2mm}} r @{\extracolsep{4mm}}
  r @{\extracolsep{-0.5mm}} r @{\extracolsep{5mm}}
  p{2.7mm} @{\extracolsep{-0.5mm}} p{2.7mm} @{\extracolsep{-0.5mm}} 
  p{2.7mm} @{\extracolsep{-0.5mm}} p{2.7mm} @{\extracolsep{-0.5mm}} 
  p{2.7mm} @{\extracolsep{-0.5mm}} p{2.7mm} @{\extracolsep{-0.5mm}} 
 c }
\hline 
\vspace{-1mm} \\
 WKK & Other & \ \ \ \ R.A. & \ \ \ Dec.& gal $\ell$ \ & gal $b \ $ 
 & 
 \multicolumn{2}{c}{D x d}& $B_{J}$ & 
 \multicolumn{5}{c}{Type} & \multicolumn{2}{c}{v$_{abs}$}  
 & \multicolumn{2}{c}{v$_{em}$} & \multicolumn{6}{l}{Identified} &  Notes \\
 Ident & Ident &(h\,\, m\,\, s) & ($\deg$\,\, $\arcmin$\,\,
 $\arcsec$) & ($\deg$) \ &($\deg$) \ 
  & \multicolumn{2}{c}{($\arcsec$)} & ($^{m}$) &
  \multicolumn{5}{c}{class.} & \multicolumn{2}{c}{km s$^{-1}$} &
  \multicolumn{2}{c}{km s$^{-1}$} & \multicolumn{6}{l}{Em. lines} & \\
\hline 
 \end{tabular*}
 \normalsize
 \end{table*}
\addtocounter{table}{-1}
\newpage

\begin{table*}
 \normalsize
 \renewcommand{\baselinestretch}{0.75}
\caption{Continued -- Redshifts of partially obscured galaxies in the Great Attractor region obtained at the SAAO.}
\scriptsize
\begin{tabular*}{18cm}{ 
  r  @{\extracolsep{2mm}} p{1.5cm} @{\extracolsep{2mm}} 
  l@{\extracolsep{3mm}} l @{\extracolsep{4mm}}
  r @{\extracolsep{2mm}}r @{\extracolsep{2mm}} 
  p{6.5mm} @{\extracolsep{0mm}} p{4.5mm} @{\extracolsep{2.5mm}}
  r @{\extracolsep{3mm}} 
  p{2.7mm} @{\extracolsep{-1mm}} p{2.7mm} @{\extracolsep{-1mm}} 
  p{2.7mm} @{\extracolsep{-1mm}} 
  p{2.7mm} @{\extracolsep{-0.5mm}} p{2.7mm} @{\extracolsep{2mm}} 
 r @{\extracolsep{2mm}} r @{\extracolsep{4mm}}
  r @{\extracolsep{-0.5mm}} r @{\extracolsep{5mm}}
  p{2.7mm} @{\extracolsep{-0.5mm}} p{2.7mm} @{\extracolsep{-0.5mm}} 
  p{2.7mm} @{\extracolsep{-0.5mm}} p{2.7mm} @{\extracolsep{-0.5mm}} 
  p{2.7mm} @{\extracolsep{-0.5mm}} p{2.7mm} @{\extracolsep{-0.5mm}} 
 c }
\hline 
\vspace{-1mm} \\
 WKK & Other & \ \ \ \ R.A. & \ \ \ Dec.& gal $\ell$ \ & gal $b \ $ 
 & 
 \multicolumn{2}{c}{D x d}& $B_{J}$ & 
 \multicolumn{5}{c}{Type} & \multicolumn{2}{c}{v$_{abs}$}  
 & \multicolumn{2}{c}{v$_{em}$} & \multicolumn{6}{l}{Identified} &  Notes \\
 Ident & Ident &(h\,\, m\,\, s) & ($\deg$\,\, $\arcmin$\,\,
 $\arcsec$) & ($\deg$) \ &($\deg$) \ 
  & \multicolumn{2}{c}{($\arcsec$)} & ($^{m}$) &
  \multicolumn{5}{c}{class.} & \multicolumn{2}{c}{km s$^{-1}$} &
  \multicolumn{2}{c}{km s$^{-1}$} & \multicolumn{6}{l}{Em. lines} & \\
\hline 
 \end{tabular*}
 \normalsize
 \end{table*}
\addtocounter{table}{-1}
\newpage

\begin{table*}
 \normalsize
 \renewcommand{\baselinestretch}{0.75}
\caption{Continued -- Redshifts of partially obscured galaxies in the Great Attractor region obtained at the SAAO.}
\scriptsize
\begin{tabular*}{18cm}{ 
  r  @{\extracolsep{2mm}} p{1.5cm} @{\extracolsep{2mm}} 
  l@{\extracolsep{3mm}} l @{\extracolsep{4mm}}
  r @{\extracolsep{2mm}}r @{\extracolsep{2mm}} 
  p{6.5mm} @{\extracolsep{0mm}} p{4.5mm} @{\extracolsep{2.5mm}}
  r @{\extracolsep{3mm}} 
  p{2.7mm} @{\extracolsep{-1mm}} p{2.7mm} @{\extracolsep{-1mm}} 
  p{2.7mm} @{\extracolsep{-1mm}} 
  p{2.7mm} @{\extracolsep{-0.5mm}} p{2.7mm} @{\extracolsep{2mm}} 
 r @{\extracolsep{2mm}} r @{\extracolsep{4mm}}
  r @{\extracolsep{-0.5mm}} r @{\extracolsep{5mm}}
  p{2.7mm} @{\extracolsep{-0.5mm}} p{2.7mm} @{\extracolsep{-0.5mm}} 
  p{2.7mm} @{\extracolsep{-0.5mm}} p{2.7mm} @{\extracolsep{-0.5mm}} 
  p{2.7mm} @{\extracolsep{-0.5mm}} p{2.7mm} @{\extracolsep{-0.5mm}} 
 c }
\hline 
\vspace{-1mm} \\
 WKK & Other & \ \ \ \ R.A. & \ \ \ Dec.& gal $\ell$ \ & gal $b \ $ 
 & 
 \multicolumn{2}{c}{D x d}& $B_{J}$ & 
 \multicolumn{5}{c}{Type} & \multicolumn{2}{c}{v$_{abs}$}  
 & \multicolumn{2}{c}{v$_{em}$} & \multicolumn{6}{l}{Identified} &  Notes \\
 Ident & Ident &(h\,\, m\,\, s) & ($\deg$\,\, $\arcmin$\,\,
 $\arcsec$) & ($\deg$) \ &($\deg$) \ 
  & \multicolumn{2}{c}{($\arcsec$)} & ($^{m}$) &
  \multicolumn{5}{c}{class.} & \multicolumn{2}{c}{km s$^{-1}$} &
  \multicolumn{2}{c}{km s$^{-1}$} & \multicolumn{6}{l}{Em. lines} & \\
\hline 
 \end{tabular*}
 \normalsize
 \end{table*}
\addtocounter{table}{-1}
\clearpage
\newpage

\begin{table*}[h]
 \normalsize
 \renewcommand{\baselinestretch}{0.75}
\caption{Continued -- Redshifts of partially obscured galaxies in the Great Attractor region obtained at the SAAO.}
\scriptsize
\begin{tabular*}{18cm}{ 
  r  @{\extracolsep{2mm}} p{1.5cm} @{\extracolsep{2mm}} 
  l@{\extracolsep{3mm}} l @{\extracolsep{4mm}}
  r @{\extracolsep{2mm}}r @{\extracolsep{2mm}} 
  p{6.5mm} @{\extracolsep{0mm}} p{4.5mm} @{\extracolsep{2.5mm}}
  r @{\extracolsep{3mm}} 
  p{2.7mm} @{\extracolsep{-1mm}} p{2.7mm} @{\extracolsep{-1mm}} 
  p{2.7mm} @{\extracolsep{-1mm}} 
  p{2.7mm} @{\extracolsep{-0.5mm}} p{2.7mm} @{\extracolsep{2mm}} 
 r @{\extracolsep{2mm}} r @{\extracolsep{4mm}}
  r @{\extracolsep{-0.5mm}} r @{\extracolsep{5mm}}
  p{2.7mm} @{\extracolsep{-0.5mm}} p{2.7mm} @{\extracolsep{-0.5mm}} 
  p{2.7mm} @{\extracolsep{-0.5mm}} p{2.7mm} @{\extracolsep{-0.5mm}} 
  p{2.7mm} @{\extracolsep{-0.5mm}} p{2.7mm} @{\extracolsep{-0.5mm}} 
 c }
\hline 
\vspace{-1mm} \\
 WKK & Other & \ \ \ \ R.A. & \ \ \ Dec.& gal $\ell$ \ & gal $b \ $ 
 & 
 \multicolumn{2}{c}{D x d}& $B_{J}$ & 
 \multicolumn{5}{c}{Type} & \multicolumn{2}{c}{v$_{abs}$}  
 & \multicolumn{2}{c}{v$_{em}$} & \multicolumn{6}{l}{Identified} &  Notes \\
 Ident & Ident &(h\,\, m\,\, s) & ($\deg$\,\, $\arcmin$\,\,
 $\arcsec$) & ($\deg$) \ &($\deg$) \ 
  & \multicolumn{2}{c}{($\arcsec$)} & ($^{m}$) &
  \multicolumn{5}{c}{class.} & \multicolumn{2}{c}{km s$^{-1}$} &
  \multicolumn{2}{c}{km s$^{-1}$} & \multicolumn{6}{l}{Em. lines} & \\
\hline
 \end{tabular*}
 \normalsize
\end{table*}

\begin{table*}
 \normalsize
 \renewcommand{\baselinestretch}{0.75}
\caption{Partially obscured galaxies in the Great Attractor region for which no redshift could be obtained.}
\label{tab2}
\scriptsize
\begin{tabular*}{15cm}{ 
  r @{\extracolsep{2mm}} p{1.5cm} @{\extracolsep{2mm}} 
  l@{\extracolsep{3mm}} l @{\extracolsep{4mm}}
  r @{\extracolsep{2mm}}r @{\extracolsep{2mm}} 
  p{6.5mm} @{\extracolsep{0mm}} p{4.5mm} @{\extracolsep{2.5mm}}
  r @{\extracolsep{3mm}} 
  p{2.7mm} @{\extracolsep{-1mm}} p{2.7mm} @{\extracolsep{-1mm}} 
  p{2.7mm} @{\extracolsep{-1mm}} p{2.7mm} @{\extracolsep{-1mm}} 
  p{2.7mm} @{\extracolsep{-0.5mm}} p{2.7mm} @{\extracolsep{2mm}} 
 l }
\hline 
\vspace{-1mm} \\
 WKK & Other & \ \ \ \ R.A. & \ \ \ Dec.& gal $\ell$ \ & gal $b \ $ 
 & 
 \multicolumn{2}{c}{D x d}& $B_{J}$ & 
 \multicolumn{6}{c}{Type} & Notes \\ 
 Ident & Ident &(h\,\, m\,\, s) & ($\deg$\,\, $\arcmin$\,\,
 $\arcsec$) & ($\deg$) \ &($\deg$) \ 
  & \multicolumn{2}{c}{($\arcsec$)} & ($^{m}$) &
  \multicolumn{6}{c}{class.} &  \\
\hline
 \end{tabular*}
 \normalsize
\end{table*}
\clearpage
\newpage

\begin{table*}
 \normalsize
 \renewcommand{\baselinestretch}{0.75}
\caption{Partially obscured galaxies in the Great Attractor region that were observed before.}
\label{tab3}
\scriptsize
\begin{tabular*}{15cm}{ 
  r @{\extracolsep{2mm}} p{1.5cm} @{\extracolsep{2mm}} 
  l@{\extracolsep{3mm}} l @{\extracolsep{4mm}}
  r @{\extracolsep{2mm}}r @{\extracolsep{2mm}} 
  p{6.5mm} @{\extracolsep{0mm}} p{4.5mm} @{\extracolsep{2.5mm}}
  r @{\extracolsep{3mm}} 
  p{2.7mm} @{\extracolsep{-1mm}} p{2.7mm} @{\extracolsep{-1mm}} 
  p{2.7mm} @{\extracolsep{-1mm}}
  p{2.7mm} @{\extracolsep{-0.5mm}} p{2.7mm} @{\extracolsep{2mm}} 
 r @{\extracolsep{2mm}} r @{\extracolsep{4mm}}
 r  }
\hline 
\vspace{-1mm} \\
 WKK & Other & \ \ \ \ R.A. & \ \ \ Dec.& gal $\ell$ \ & gal $b \ $ 
 & 
 \multicolumn{2}{c}{D x d}& $B_{J}$ & 
 \multicolumn{5}{c}{Type} & \multicolumn{2}{c}{v$_{lit}$} & Notes \\ 
 Ident & Ident &(h\,\, m\,\, s) & ($\deg$\,\, $\arcmin$\,\,
 $\arcsec$) & ($\deg$) \ &($\deg$) \ 
  & \multicolumn{2}{c}{($\arcsec$)} & ($^{m}$) &
  \multicolumn{5}{c}{class.} &  \multicolumn{2}{c}{km s$^{-1}$} & \\
\hline
 \end{tabular*}
 \normalsize
\end{table*}
\clearpage
\newpage

%% file: H1430tab4.tex
\begin{table*}[t]
 \normalsize
 \renewcommand{\baselinestretch}{0.75}
\caption{Central velocity dispersion measurements for 34 galaxies in the Norma cluster.}
\label{tab4}
\scriptsize
\begin{tabular*}{17.5cm}{ 
  r @{\extracolsep{2mm}} p{1.5cm} @{\extracolsep{2mm}} 
  l@{\extracolsep{3mm}} l @{\extracolsep{4mm}}
  r @{\extracolsep{2mm}}r @{\extracolsep{2mm}} 
  p{6.5mm} @{\extracolsep{0mm}} p{4.5mm} @{\extracolsep{2.5mm}}
  r @{\extracolsep{3mm}} r @{\extracolsep{3mm}} 
  p{2.7mm} @{\extracolsep{-1mm}} p{2.7mm} @{\extracolsep{-1mm}} 
  p{2.7mm} @{\extracolsep{-1mm}} 
  p{2.7mm} @{\extracolsep{-0.5mm}} p{2.7mm} @{\extracolsep{2mm}} 
 r @{\extracolsep{2mm}} r @{\extracolsep{2mm}}
  r @{\extracolsep{2mm}} r @{\extracolsep{3mm}} r @{\extracolsep{1mm}}  }
\hline 
\vspace{-1mm} \\
 WKK & Other & \ \ \ \ R.A. & \ \ \ Dec.& gal $\ell$ \ & gal $b \ $ 
 & 
 \multicolumn{2}{c}{D x d}& $B_{J}$ & E(B-V) &
 \multicolumn{5}{c}{Type} & \multicolumn{2}{c}{v$_{abs}$}  
 & $\sigma_{raw}$ & \multicolumn{2}{c}{{\rm log}({$\sigma_{o})$}} \\
 Ident & Ident &(h\,\, m\,\, s) & ($\deg$\,\, $\arcmin$\,\,
 $\arcsec$) & ($\deg$) \ &($\deg$) \ 
  & \multicolumn{2}{c}{($\arcsec$)} & ($^{m}$) &  &
  \multicolumn{5}{c}{class.} & \multicolumn{2}{c}{km s$^{-1}$} &
  km s$^{-1}$ & \multicolumn{2}{c}{}   \\
\hline
 \end{tabular*}
 \normalsize
\end{table*}